\documentclass[12pt,a4paper]{article}

\pdfoutput=1

\usepackage[DIV13,BCOR0mm]{typearea}

\usepackage{natbib}
\bibpunct[, ]{(}{)}{,}{a}{}{,}

\usepackage{graphicx}

\newcommand{\beq}{\begin{equation}}
\newcommand{\eeq}{\end{equation}}
\newcommand{\beqa}{\begin{eqnarray}}
\newcommand{\eeqa}{\end{eqnarray}}
\newcommand{\beqan}{\begin{eqnarray*}}
\newcommand{\eeqan}{\end{eqnarray*}}

\newcommand{\E}{\mathrm{e}}

\newcommand{\degC}{\ensuremath{^\circ\mathrm{C}}}    % degrees C
\def\pmb#1{\setbox0=\hbox{#1}%
  \kern-.025em\copy0\kern-\wd0
  \kern.05em\copy0\kern-\wd0
  \kern-.025em\raise.0433em\box0}

\def\pmbm#1{\setbox0=\hbox{$#1$}%
  \kern-.025em\copy0\kern-\wd0
  \kern.05em\copy0\kern-\wd0
  \kern-.025em\raise.0433em\box0}

\def\vecbi#1{\relax\ifmmode\mathchoice
  {\mbox{\boldmath$\relax\displaystyle#1$}}
  {\mbox{\boldmath$\relax\textstyle#1$}}
  {\mbox{\boldmath$\relax\scriptstyle#1$}}
  {\mbox{\boldmath$\relax\scriptscriptstyle#1$}}\else
  \hbox{\boldmath$\relax\textstyle#1$}\fi} % boldface italic symbols

\def\vecbu#1{\relax\ifmmode\mathchoice
  {\mbox{\boldmath$\bf\displaystyle#1$}}
  {\mbox{\boldmath$\bf\textstyle#1$}}
  {\mbox{\boldmath$\bf\scriptstyle#1$}}
  {\mbox{\boldmath$\bf\scriptscriptstyle#1$}}\else
  \hbox{\boldmath$\bf\textstyle#1$}\fi}    % boldface upright symbols

\def\tenssu#1{\relax\ifmmode\mathchoice
    {\mbox{$\sf\displaystyle#1$}}
    {\mbox{$\sf\textstyle#1$}}
    {\mbox{$\sf\scriptstyle#1$}}
    {\mbox{$\sf\scriptscriptstyle#1$}}\else
    \hbox{$\sf\textstyle#1$}\fi}           % sans-serif upright symbols

\begin{document}

\title{\textbf{Decay of the Greenland Ice Sheet due to surface-meltwater-induced
               acceleration \\{}of basal sliding}}

\author{\textsc{Ralf Greve}\thanks{E-mail: greve@lowtem.hokudai.ac.jp}\\
        \textsc{Shin Sugiyama}\\[0.5ex]
        {\normalsize Institute of Low Temperature Science, Hokkaido University,}\\[-0.25ex]
        {\normalsize Kita-19, Nishi-8, Kita-ku, Sapporo 060-0819, Japan}}

\date{}

\maketitle

\begin{abstract}

Simulations of the Greenland Ice Sheet are carried out with a
high-resolution version of the ice-sheet model SICOPOLIS for
several global-warming scenarios for the period 1990--2350. In
particular, the impact of surface-meltwater-induced
acceleration of basal sliding on the stability of the ice sheet
is investigated. A parameterization for the acceleration effect
is developed for which modelled and measured mass losses of the
ice sheet in the early 21st century agree well. The main
findings of the simulations are: (i) the ice sheet is generally
very susceptible to global warming on time-scales of centuries,
(ii) surface-meltwater-induced acceleration of basal sliding
leads to a pronounced speed-up of ice streams and outlet
glaciers, and (iii) this ice-dynamical effect accelerates the
decay of the Greenland Ice Sheet as a whole significantly, but
not catastrophically, in the 21st century and beyond.

\end{abstract}

\section{Introduction}
\label{sect_intro}

In Chapter 10 (``Global Climate Projections'') of the Fourth
Assessment Report (AR4) of the United Nations Intergovernmental
Panel on Climate Change (IPCC), an increase of the mean global
sea level by 18--59~cm for the 21st century (more precisely:
2090--2099 relative to 1980--1999) is projected for the six
SRES marker scenarios B1, B2, A1B, A1T, A2 and A1FI (Meehl
et~al.\ \citeyear{meehl_etal_07_short}). The main causes for
this sea level rise are thermal expansion of sea water and
melting of glaciers and small ice caps, and to a lesser extent
changes of the surface mass balance of the Greenland and
Antarctic Ice Sheets. However, recent observations suggest that
ice flow dynamics could lead to additional sea level rise, and
this problem is explicitly stated in the AR4:
\begin{quote}
\emph{``Dynamical processes related to ice flow not included in
current models but suggested by recent observations could
increase the vulnerability of the ice sheets to warming,
increasing future sea level rise. Understanding of these
processes is limited and there is no consensus on their
magnitude.''} \citep{ipcc_07a_short}.
\end{quote}
These conjectured dynamical processes are (i) basal sliding
accelerated by surface meltwater, (ii) reduced buttressing due
to the loss of ice shelves, and (iii) penetration of ocean
water under the ice. The first process is more relevant for the
Greenland Ice Sheet (the focus of this study), whereas the
latter two may affect the stability of the West Antarctic Ice
Sheet. On the observational side, recent results from satellite
gravity measurements for the period 2002-2005 indicate
surprisingly large mass losses of
$239\pm{}23\;\mathrm{km^3\,a^{-1}}$
($0.66\pm{}0.06\;\mathrm{mm\;a^{-1}}$ sea level equivalent) for
the Greenland Ice Sheet \citep{chen_etal_06}. Furthermore,
major outlet glaciers (Jacobshavn ice stream, Kangerdlugssuaq
and Helheim glaciers) have sped up drastically since the 1990's
\citep{rignot_kanagaratnam_06}.

\section{Ice-sheet model SICOPOLIS}
\label{sect_modelling}

For this study, we use the ice-sheet model SICOPOLIS
(``SImulation COde for POLythermal Ice Sheets''), which
simulates the large-scale dynamics and thermodynamics (ice
extent, thickness, velocity, temperature, water content and
age) of ice sheets three-di\-men\-sio\-nal\-ly and as a
function of time (\citeauthor{greve_97b} \citeyear{greve_97b};
\texttt{http://sicopolis.greveweb.net/}). It is based on the
shallow-ice approximation \citep[e.g.][]{hutter_83} and the
rheology of an incompressible, heat-conducting power-law fluid
[Glen's flow law, see \citet{paterson_94}]. External forcing is
specified by (i) the mean annual air temperature at the ice
surface, (ii) the surface mass balance (precipitation minus
runoff), (iii) the sea level surrounding the ice sheet and (iv)
the geothermal heat flux prescribed at the bottom of a
lithospheric thermal boundary layer of 5~km thickness. For all
simulations of this study, the horizontal resolution is 10~km,
the vertical resolution is 81 grid points for the cold-ice
column, 11 grid points for the basal layer of temperate ice (if
existing) and 11 grid points for the lithosphere layer, the
time-step is 0.25~a, and the geothermal heat-flux distribution
and model parameters are those used by \citet{greve_05a}.

\section{WRE1000 scenario}
\label{sect_wre1000}

\begin{figure}[htb]
  \centering
  \includegraphics[width=0.55\textwidth, clip]{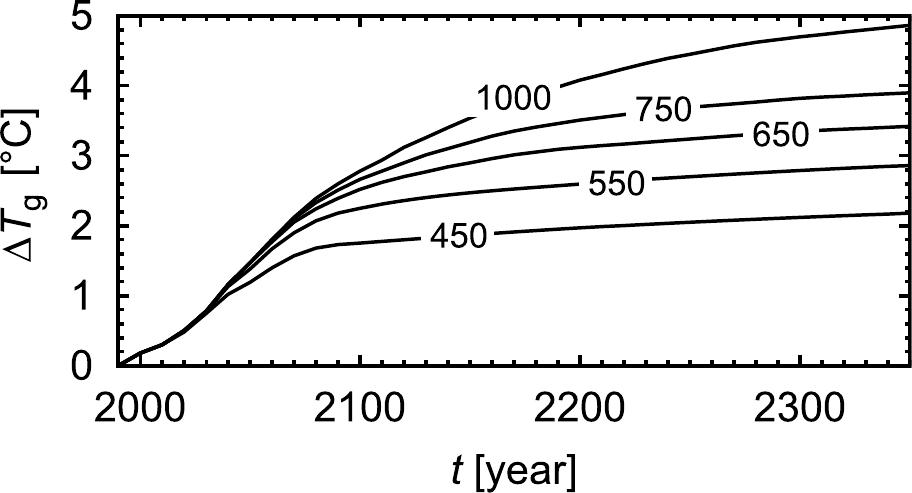}
  \caption{Global mean temperature change $\Delta{}T_\mathrm{g}$
  for the scenarios WRE450, WRE550, WRE650, WRE750 and WRE1000
  (stabilisation scenarios for atmospheric CO$_2$),
  by Cubasch et~al.\ (\citeyear{cubasch_etal_01_short}).
  In this study, only the WRE1000 scenario will be considered.}
  \label{fig_wre_temp}
\end{figure}

Future global warming shall be prescribed
exemplarily by the WRE1000 scenario, which assumes
stabilisation of the atmospheric CO$_2$ concentration at
1000~ppm (Cubasch et~al.\ \citeyear{cubasch_etal_01_short}).
The corresponding temperature change from 1990 (the
``present'') until 2350 is shown in Fig.~\ref{fig_wre_temp}
(along with similar scenarios with lower stabilisation
concentrations).

In order to obtain temperature and precipitation forcings for
the Greenland Ice Sheet, the argumentation by \citet{greve_04}
is followed. The surface temperatures shown in
Fig.~\ref{fig_wre_temp} are amplified by a factor 2 and imposed
as uniform increases over the ice sheet, and the precipitations
are assumed to increase by $5\%$ per degree of
ice-sheet-surface-temperature change. Surface melting is
parameterized by the degree-day method in the version by
\citet{greve_05a}. This approach is a critical simplification,
and it should rather be replaced by an energy-balance model for
more accurate results. However, since the objective of this
study is to assess the impact of ice-dynamical processes on the
decay of the Greenland Ice Sheet rather than making precise
predictions of the decay itself, the use of the degree-day
method is a reasonable compromise.

\section{Basal sliding}
\label{sect_basal_sliding}

Basal sliding is described by a Weertman-type
sliding law in the form of \citet{greve_otsu_07}, based on
\citet{greve_etal_98} and modified to allow for sub-melt
sliding \citep{hindmarsh_lemeur_01},
\beq
  v_\mathrm{b} = -C_\mathrm{b}\,\E^{T'_\mathrm{b}/\gamma_\mathrm{sms}}
                 \times \frac{\tau_\mathrm{b}^3}{P_\mathrm{b}^2}\,,
  \label{eq_vb_ref}
\eeq
where $v_\mathrm{b}$ is the basal-sliding velocity,
$C_\mathrm{b}$ the sliding coefficient, $\tau_\mathrm{b}$ the
basal shear traction in the bed plane, $\rho$ the ice density,
$g$ the gravity acceleration, $H$ the ice thickness and
$P_\mathrm{b}{=}\rho{}gH$ the overburden pressure. The term
$e^{T'_\mathrm{b}/\gamma_\mathrm{sms}}$ represents the
exponentially diminishing sub-melt sliding, where
$T'_\mathrm{b}$ is the temperature relative to pressure melting
(in $\degC$) and $\gamma_\mathrm{sms}{=}1\degC$ the
sub-melt-sliding coefficient.

Acceleration of basal sliding by surface meltwater is
parameterized by an extension of the approach by
\citet{greve_otsu_07}. The sliding coefficient is expressed as
\beq
  C_\mathrm{b} =
  C_\mathrm{b}^0\,\left(1+\frac{\gamma}{H^r}M^s\right)\,,
  \label{eq_basal_sliding_smw}
\eeq
where $C_\mathrm{b}^0=11.2\,\mathrm{m\,a^{-1}\,Pa^{-1}}$, $M$
is the surface melt rate (runoff), $\gamma$ is the surface
meltwater coefficient, and $r$ and $s$ are adjustable
exponents. The idea behind this parameterization is to relate
the sliding speed-up to the local surface melt rate, and
account for the less efficient percolation of meltwater to the
base in regions where the ice is thick by the dependency on the
inverse ice thickness.

The parameterization employed by \citet{greve_otsu_07}
corresponds to $(r,s)=(0,1)$. For this case, the authors show
in their Appendix~A that data reported by
\citet{zwally_etal_02} from the Swiss Camp in central west
Greenland give rise to the estimate
$\gamma=0.1\;\mathrm{a\;m^{-1}}$. We will also consider the
cases $(r,s)=(0,2)$ and $(1,1)$, for which the same arguments
lead to estimates of $\gamma=0.05\;\mathrm{a^2\;m^{-2}}$ and
$100\;\mathrm{a}$, respectively.

\section{Simulations}
\label{sect_simulations}

\subsection{Set-up}
\label{ssect_setup}

Five simulations with different settings for the acceleration
of basal sliding by surface meltwater will be discussed in
order to investigate to what extent this process can increase
the vulnerability of the Greenland Ice Sheet to future warming.
In run~\#1, acceleration of basal sliding by surface meltwater
is not considered ($\gamma=0$). Runs~\#2-4 have been conducted
with $(r,s)=(0,1)$, $(0,2)$ and $(1,1)$, respectively, and
values of $\gamma$ chosen according to the estimates given at
the end of Sect.~\ref{sect_basal_sliding} (designated in
Table~\ref{tab_results} as ``100\%''). Run \#5 corresponds to
the most extreme scenario considered by \citet{greve_otsu_07},
with the settings $(r,s)=(0,1)$ and
$\gamma=5\;\mathrm{a\;m^{-1}}$ (50 times the above estimate,
therefore designated in Table~\ref{tab_results} as ``5000\%'').
All simulations start with the present-day ice sheet as initial
condition, and the model time is from 1990 until 2350.

\begin{table}[htb]
  \begin{center}\begin{tabular}{ccccccc} \hline
  Run\rule{0ex}{2.7ex}
  & $\gamma$
  & $(r,s)$
  & $\dot{V}_\mathrm{2002-2005}$
  & $\Delta{}V_\mathrm{2100}$
  & $\Delta{}V_\mathrm{2200}$
  & $\Delta{}V_\mathrm{2300}$ \\
  & & & $[\mathrm{km^3\,a^{-1}}]$ & $[\mathrm{m\;SLE}]$
      & $[\mathrm{m\;SLE}]$ & $[\mathrm{m\;SLE}]$\rule[-1.2ex]{0ex}{1.2ex} \\ \hline
  \#1 & $\,0$ & --- &        $\;\;37.7$ & $0.12$ & $0.55$ & $1.21$\rule{0ex}{2.4ex} \\
  \#2 & $100\%$ & $(0,1)$ & $111.6$ & $0.14$ & $0.60$ & $1.31$ \\
  \#3 & $100\%$ & $(0,2)$ & $172.0$ & $0.17$ & $0.68$ & $1.48$ \\
  \#4 & $100\%$ & $(1,1)$ & $248.9$ & $0.18$ & $0.67$ & $1.42$ \\
  \#5 & $5000\%\;\;$ & $(0,1)$ &      $1627.8\;\,$ & $0.58$ & $1.51$ & $2.71$ \\ \hline
  \end{tabular}\end{center}
  \caption{Set-up and results of runs \#1-5.
  For the meaning of the basal-sliding parameters $\gamma$,
  $r$, $s$ see Eq.~(\ref{eq_basal_sliding_smw}) and the accompanying
  text. $\dot{V}_\mathrm{2002-2005}$ denotes the average
  loss of ice volume between 2002 and 2005, whereas
  $\Delta{}V_\mathrm{2100}$, $\Delta{}V_\mathrm{2200}$ and $\Delta{}V_\mathrm{2300}$
  are the losses of ice volume by 2100, 2200 and 2300, respectively,
  compared to 1990. The latter are expressed in meters of sea-level equivalent.}
  \vspace*{-1ex}
  \label{tab_results}
\end{table}

\subsection{Results}
\label{ssect_results}

An overview of the main results is given in
Table~\ref{tab_results}. The average loss of ice volume between
2002 and 2005 can be compared with the measured value by
\citet{chen_etal_06} of $239\pm{}23\;\mathrm{km^3\,a^{-1}}$
(see introduction), thus providing an observational constraint
for the simulations. Evidently, the ice-volume loss is far too
small for run~\#1 (no acceleration of basal sliding by surface
meltwater), and it is too small by about a factor 2 for run~\#2
[$(r,s)=(0,1)$]. By contrast, the agreement is quite good for
run~\#3 [$(r,s)=(0,2)$] and very good for run~\#4
[$(r,s)=(1,1)$]. On the other hand, the extreme case of run~\#5
[$(r,s)=(0,1)$, very large $\gamma$] produces more than 6 times
more ice-volume loss than observed. Therefore, runs ~\#3 and 4
seem to be most realistic.

From a theoretical point of view, the set-up of run~\#4 is
preferable to that of run~\#3, because it is clear that the
percolation of surface meltwater to the base will be the less
efficient the thicker the ice is. This is accounted for in
run~\#4, for which the acceleration of basal sliding decreases
with increasing ice thickness ($r=1$), whereas this is not the
case in run~\#3 ($r=0$). Consequently, run~\#4 shall be
considered as the ``best'' simulation.

\begin{figure}[p]
  \centering
  \includegraphics[scale=1.0, clip]{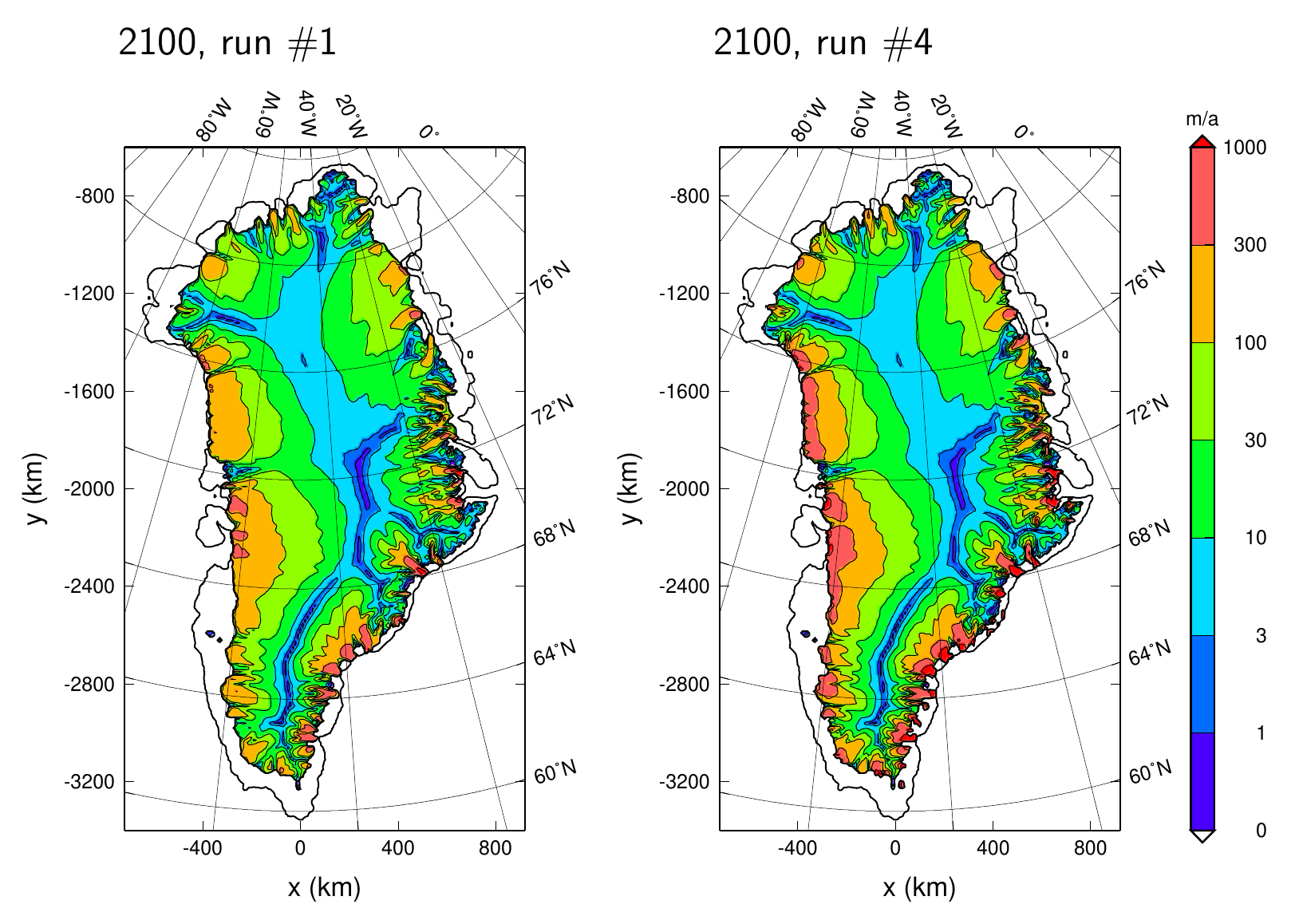}
  \caption{Simulated surface velocity of the Greenland Ice Sheet.
  Left: Run \#1
  (no acceleration of basal sliding by surface meltwater), year 2100.
  Right: Run \#4
  (``best'' simulation, including acceleration of basal sliding by surface
  meltwater), year 2100.
  Both simulations reproduce the organization of the drainage pattern
  into ice streams and outlet glaciers, despite the use of the shallow-ice
  approximation and the large-scale (10~km) resolution.
  The dynamic acceleration effect in run~\#4 is clearly visible all around
  the ice margin and leads to faster decay.}
  \label{fig_surf_vel}
\end{figure}

Comparison of the results of run~\#4 and run~\#1 (no
acceleration of basal sliding by surface meltwater) shows that
the contribution to sea-level rise by 2100 is $\sim{}50\%$
larger for run~\#4 (0.18 vs.\ 0.12~m). The impact of the
acceleration effect on ice flow becomes evident by inspection
of Fig.~\ref{fig_surf_vel} which shows the simulated surface
velocities in 2100 for the two runs. Therefore, the
acceleration of basal sliding by surface meltwater, which is
most likely the major ice-dynamical process relevant for the
Greenland Ice Sheet in the context of global warming, has a
significant, but not catastrophic effect on the decay of the
ice sheet in the 21st century.

\begin{figure}[p]
  \centering
  \includegraphics[scale=0.93, clip]{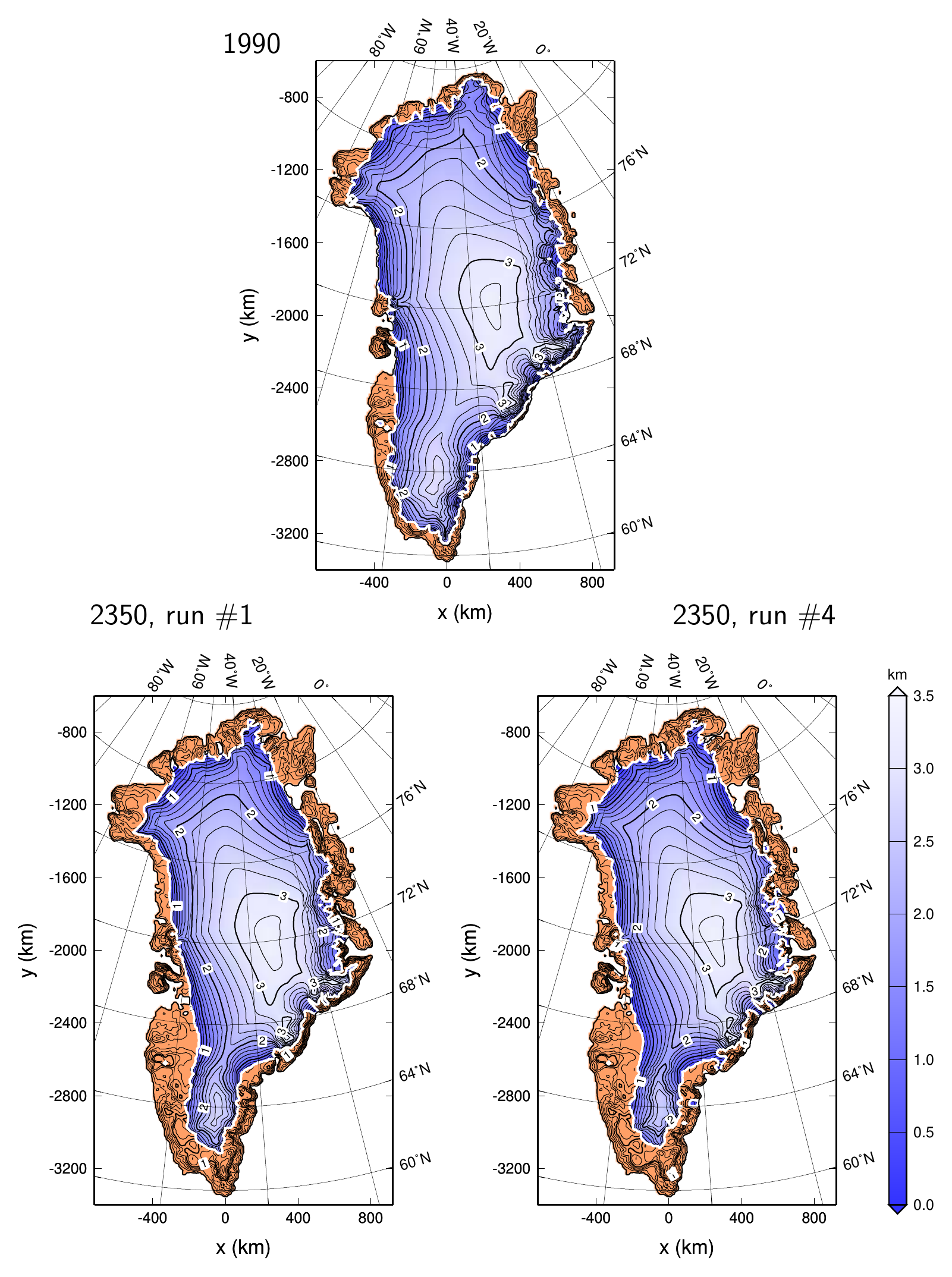}
  \caption{Simulated surface topography of the Greenland Ice Sheet.
  Top: Initial condition for the year 1990.
  Bottom left: Run \#1
  (no acceleration of basal sliding by surface meltwater), year 2350.
  Bottom right: Run \#4
  (``best'' simulation, including acceleration of basal sliding by surface
  meltwater), year 2350.
  The difference between 1990 and 2350 amounts to $1.59\;\mathrm{m\;SLE}$
  for run~\#1 and to $1.84\;\mathrm{m\;SLE}$ for run~\#4.}
  \label{fig_surf_topo}
\end{figure}

The absolute difference between the two runs becomes larger in
the more distant future; however, the relative difference
becomes smaller: by 2200 the contribution to sea-level rise is
0.12~m ($\sim{}22\%$) larger for run~\#4, and by 2300 it is
0.21~m ($\sim{}17\%$) larger. Figure~\ref{fig_surf_topo} shows
the simulated surface topographies in 2350 (at the end of the
simulations). It is nicely illustrated that for both runs ~\#1
and \#4 the ice sheet shows a strong response on the imposed
warming scenario and retreats all around the margin (most
pronounced in the south-west), while the
surface-meltwater-induced acceleration of basal sliding
accounted for in run ~\#4 speeds up the decay.

Two additional simulations with larger exponents, namely
$(r,s)=(1,2)$ and $(2,1)$, and values of $\gamma$ chosen in
analogy to the ``100\%'' runs~\#2-4, have also been conducted.
For these cases, maximum surface velocities of more than
$100\;\mathrm{km\,a^{-1}}$ occur close to the ice margin, which
is unrealistic. Apparently, the speed-up effect is too
pronounced for these settings, and so they have been discarded.

\clearpage

\clearpage

\clearpage

\section{Conclusion}
\label{sect_conclusion}

The simulations discussed in this study suggest that
ice-dynamical processes can speed up the decay of the Greenland
Ice Sheet significantly in the 21st century and beyond.
However, a catastrophically accelerated decay can only be
obtained with unrealistic parameter settings and thus seems to
be unlikely.

\section*{Acknowledgements}

% The authors wish to thank J.~Church...

This study was supported by a Grant-in-Aid for
Scientific Research (Category B, No.\ 18340135) from the Japan
Society for the Promotion of Science.

% \section*{References}

% \bibliographystyle{ralf}
% \bibliography{journals2_ralf,publ_ralf}

\end{document}